\documentclass[amsmath,amssymb,superscriptaddress, prl,twocolumn]{revtex4-1}

\usepackage{graphicx}
\usepackage{dcolumn}
\usepackage{bm}
\usepackage{amsmath}
\usepackage{amssymb,amsthm}
\usepackage{color}
\usepackage{epstopdf} 
\usepackage[sans]{dsfont}
\usepackage{accents}
\usepackage[utf8]{inputenc}
\usepackage[T1]{fontenc}
\usepackage[english]{babel}
\usepackage{hyperref}
\usepackage{mathbbol,bbm}
\usepackage{float}
\usepackage{accents}

\graphicspath{{Figures/}}

\newcommand\numberthis{\addtocounter{equation}{1}\tag{\theequation}}
\newcommand{\Id}{\mathbb{1}}		
\def\ket#1{\mathinner{|{#1}\rangle}}		
\newcommand{\ketbra}[2]{\left|#1\rangle\langle #2\right|}		
\newcommand{\matelem}[3]{\langle #1|#2|#3\rangle}		
\newcommand{\op}[1]{\ensuremath{\hat{#1}}}
\newcommand{\avg}[1]{\ensuremath{\overline{#1}~}}
\newcommand{\ens}[1]{\ensuremath{\left\{#1\right\}}}
\newcommand{\commutator}[2]{\left[#1,#2\right]}

\newcommand{\limit}[2]{\underset{#1\rightarrow#2}{\lim}}

\newcommand{\eqdef}{:=}

\newcommand{\Ham}{\ensuremath{\op{\textrm{H}}}}

\newcommand{\Enslabel}{\ensuremath{\lambda}}
\newcommand{\Dm}{\ensuremath{\op{\rho}}}

\newcommand{\Dmaponly}{\ensuremath{\Lambda}}
\newcommand{\Dmap}[1]{\ensuremath{\Dmaponly_t \left[#1\right]}}

\newcommand{\Dim}{\ensuremath{d}}

\newcommand{\Avgham}{\ensuremath{\avg{\Ham}}}
\newcommand{\Uop}{\ensuremath{\op{U}}}

\newcommand{\HamI}{\Ham^\textrm{SE}}

\begin{document}

\title{Open system model for quantum dynamical maps with classical noise and corresponding master equations}

\author{Chahan M. Kropf} 
\email{Chahan.Kropf@physik.uni-freiburg.de}
\affiliation{Physikalisches Institut, Albert-Ludwigs-Universit\"at Freiburg, Hermann-Herder-Str.~3, D-79104 Freiburg, Germany}

\author{Vyacheslav N. Shatokhin}
\affiliation{Physikalisches Institut, Albert-Ludwigs-Universit\"at Freiburg, Hermann-Herder-Str.~3, D-79104 Freiburg, Germany}

\author{Andreas Buchleitner}
\affiliation{Physikalisches Institut, Albert-Ludwigs-Universit\"at Freiburg, Hermann-Herder-Str.~3, D-79104 Freiburg, Germany}

\date{\today}

\begin{abstract}
\noindent
We show how random unitary dynamics arise from the coupling of an open quantum system to a static environment. Subsequently, we derive a master equation for the reduced system random unitary dynamics and study three specific cases: commuting system and interaction Hamiltonians, the short-time limit, and the Markov approximation.  
\end{abstract}

\maketitle


 In information theory, the dynamics of quantum systems subject solely to classical uncertainty can be modelled by random mixtures of unitary dynamics, or random unitary maps \cite{Bengtsson2006}, 
\begin{align}\label{eq:Rnd unit map general}
	\Dmaponly[\Dm] = \sum_\Enslabel p_\Enslabel W_\Enslabel \Dm W_\Enslabel^\dagger,
\end{align}
which are convex combinations of unitary Kraus operators $W_\Enslabel$ weighted with normalized probabilities $p_\Enslabel$,  $\sum_\Enslabel p_\Enslabel=1$, $p_\Enslabel \in [0,1]$. 

On the one hand, the Kraus form \eqref{eq:Rnd unit map general} has proven to be a powerful tool in derivations of general mathematical relations characterizing the dynamics of quantum systems subject to classical uncertainty \cite{Gregoratti2003,Audenaert2008,Buscemi2006,Emerson2005,Rosgen2008a}. 
Moreover, random unitary maps have been used to study generic properties of quantum systems, such as Markovianity \cite{Chruscinski2015}.  
However, the abstract formulation in terms of Kraus operators does not provide a direct physical description of quantum dynamics in terms of experimentally realizable Hamiltonians. Therefore, the experimental implementation of random unitary maps, e.g., a mixture of Pauli channels that are easily studied on a mathematical level \cite{Chruscinski2013}, may require complicated experimental setups \cite{Liu2011}. 

On the other hand, random unitary maps based on 
microscopic models describing disordered systems \cite{Kropf2016a,Gneiting2016} or open quantum systems \cite{Bengtsson2006,Chen2017} are physically motivated. Whereas these models, in general, give rise to mathematically rather complicated random unitary dynamics, they provide access to extensive methods developed in the theory of open quantum systems and of disordered systems to study the emerging dynamics. Among various methods, a distinguished role is played by master equations. 

It was demonstrated in \cite{Kropf2016a} how to derive and interpret master equations in the context of disordered systems. In this contribution, we will explore the open quantum system approach. To begin with, we show how random unitary dynamics arise from the coupling of an open system to a static environment. Subsequently, we derive the corresponding master equation in the Born approximation. We then discuss more explicitly three cases: commuting system and interaction Hamiltonians, the short-time limit and the semi-group Markov approximation. The first case leads to pure dephasing dynamics, the second one captures the generic Gaussian incoherent dynamics at times shorter than the Heisenberg time, and the third case is characterized by the free retarded Green's function associated with the system Hamiltonian. These findings illustrate the intrinsic connection between disordered systems, open quantum systems and random unitary maps. 

\begin{figure}
	\centering
	\includegraphics[width=0.45\textwidth]{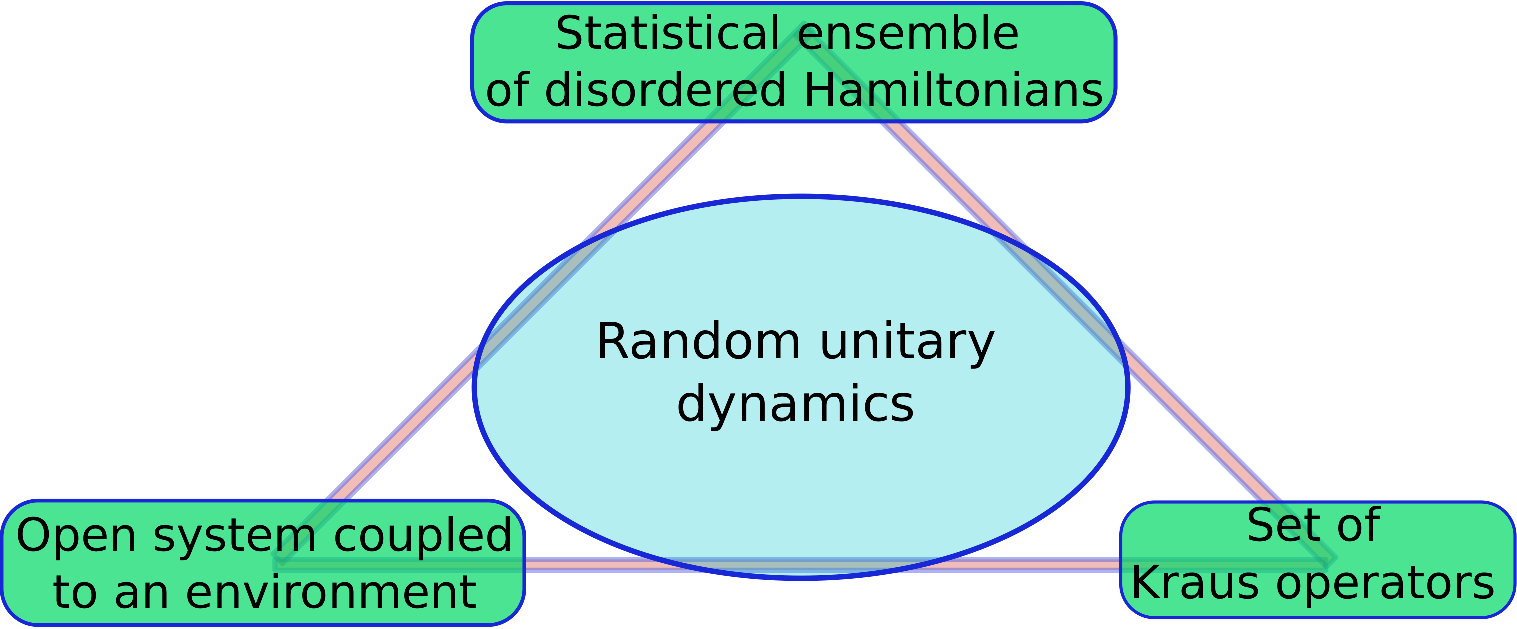}
	\caption[Open system perspective]{Random unitary dynamics can be formulated in three different frameworks. On can start from the Kraus representation, as is customary in information theory, or from an ensemble disordered time-independent Hamiltonians, as is done in disorder physics, or from the coupling of an open system to an environment, as in the theory of open quantum systems.}
	\label{fig:Equivalence illustration}
\end{figure}

~\linebreak
~

In the following, we consider a specific random unitary dynamics described by the following dynamical map:
\begin{align}\label{eq:Rndunit Dynamical map}
	\Dmaponly_t[\Dm] = \int d\Enslabel p(\Enslabel) U_\Enslabel(t) \Dm U_\Enslabel^\dagger(t),
\end{align}
where $U_\Enslabel (t) = e^{-it\Ham_\Enslabel/\hbar}$ is the unitary time-evolution operator arising from the Hamiltonian $\Ham_\Enslabel$. Note that, mathematically, it is common to consider a sum of unitary evolutions (c.f. Eq.~\eqref{eq:Rnd unit map general}). However, for physically motivated models, it is often more natural to consider continuous combinations of unitary maps. For instance, such a situation arises when an open system is embedded in an environment with a continuous density of states. Hence, we have an integral in the right hand side of Eq.~\eqref{eq:Rndunit Dynamical map}, with a normalized probability distribution $p(\Enslabel)$.

As discussed in \cite{Kropf2016a}, Eq.~\eqref{eq:Rndunit Dynamical map} can be interpreted as the ensemble-averaged dynamics of a disordered quantum system characterized by the Hamiltonian $\Ham_\Enslabel = \Avgham+ \delta\Ham_\Enslabel$, where $\delta\Ham_\Enslabel$ are random Hamiltonians distributed according to the probability distribution $p(\Enslabel)$ and $\Avgham :=\int d\Enslabel p(\Enslabel) \Ham_\Enslabel$ is the average Hamiltonian. The dynamics of a single realization of the disorder is then given by $\Dm_\Enslabel(t) = U_\Enslabel(t) \Dm(0) U_\Enslabel^\dagger(t)$, with $U_\Enslabel(t) = e^{-it(\Avgham+\delta\Ham_\Enslabel)/\hbar}$. Correspondingly, the ensemble-averaged dynamics is obtained as $\overline{\Dm}(t) := \int d\Enslabel p(\Enslabel) \rho_\Enslabel(t) = \Dmaponly_t[\Dm(0)]$, which is a random unitary channel in the form of Eq.~\eqref{eq:Rndunit Dynamical map}. The significance of this result stems from the fact that many quantum transport  and/or interaction phenomena in complex systems can effectively be described using models of disorder. Relevant examples include the electronic transport in wires with impurities  (Anderson model) \cite{Anderson1958}, the propagation of entangled photons across atmospheric turbulence \cite{Roux2014a}, the exciton transport in molecular complexes \cite{Lee2015}, and the inhomogeneous broadening of spectral line widths \cite{Slichter1990}.

 Alternatively, as we show in the following, one can also conceive the dynamics \eqref{eq:Rndunit Dynamical map} as arising from the coupling of an open quantum system to a static environment. 
The open system and its environment are characterized by the total Hamiltonian $\Ham^T = \Ham^S + \HamI + \Ham^E$, where $\Ham^S$, $\HamI$ and $\Ham^E$ are the Hamiltonians of the open system, the interaction and the environment, respectively. We assume an initial product (uncorrelated) state, $\Dm_0 = \Dm_0^S \otimes \Dm_0^E $, of the system and environment. Furthermore, we assume that the environment part $\Dm_0^E$ is a stationary state of $\Ham^E$: 
\begin{align}\label{eq:Open system initial state}
	\Dm_0 = \Dm_0^S \otimes \int d\Enslabel \, p(\Enslabel )\ketbra{\Enslabel}{\Enslabel},
\end{align}
 with the (orthonormal) eigenstates $\ket{\Enslabel}$ of $\Ham^E$, and the normalized probability density $p(\Enslabel)$, $\int d\Enslabel p(\Enslabel)=1$. 

In order to generate random unitary dynamics, we further assume that the state of the environment is time-independent. In other words, we impose the Born approximation, $\Dm(t) \approx \Dm^S(t)\otimes \Dm^E$, to hold at arbitrary times. This can be achieved by demanding 
\begin{align}\label{eq:HECommutatorHSE}
[\Ham^E, \HamI]=0.
\end{align}
 In this case, $\Ham^E$ and $\HamI$ share a common eigenbasis and, thus, we can parametrize the interaction Hamiltonian as
 \begin{align}\label{eq:Open syst Ham}
 \HamI= \int d\Enslabel \,\Ham_\Enslabel \otimes \ketbra{\Enslabel}{\Enslabel}.
 \end{align}
The simplest example of an interaction Hamiltonian in the form of Eq.~\eqref{eq:Open syst Ham} is $\HamI = \Ham^{A}\otimes\Ham^{B}$. Indeed, if we expand $\Ham^{B}$ in its eigenbasis $\ket{\Enslabel}$ (with eigenvalues $E_\Enslabel^B$) we obtain $\HamI = \int_\Enslabel E^B_\Enslabel \Ham^{A} \otimes \ketbra{\Enslabel}{\Enslabel}$. Defining $\Ham_\Enslabel \equiv E_\Enslabel \Ham^{A}$ then yields Eq.~\eqref{eq:Open syst Ham}. 
 
 Given the initial state \eqref{eq:Open system initial state} and condition \eqref{eq:HECommutatorHSE}, the Hamiltonian of the environment plays no role in the dynamics and, thus, we can restrict our further analysis to $\Ham = \Ham^S+\HamI$. This Hamiltonian has a block diagonal structure, 
\begin{align}\label{eq:HS and HSE}
	\Ham = \begin{pmatrix}
		\Ham^S+\Ham_1 & ~ & ~ & ~\\
		~ & \Ham^S+\Ham_2& ~ & ~ \\
		~ & ~ & \Ham^S+\Ham_3 & ~ \\
		~ & ~ & ~ & \ldots
	\end{pmatrix},
\end{align} 
where each block corresponds to one realisation of the random unitary channel. It is useful to include the average Hamiltonian, $\Avgham = \int_\Enslabel p(\Enslabel) \Ham_\Enslabel$, into the definition of the system Hamiltonian:
\begin{align}\label{eq:H-Transfo}
	\Ham^S \rightarrow \Ham^S + \Avgham \;\; , \;\; \Ham_\Enslabel \rightarrow \Ham_\Enslabel - \Avgham.
\end{align}
With this transformation, the Hamiltonian \eqref{eq:HS and HSE} remains unchanged, but the average $ \int d\Enslabel p(\Enslabel) \Ham_\Enslabel=0$ now vanishes.

The dynamics of the total system is then characterized by the unitary time-evolution operator $U(t) = e^{-it\Ham/\hbar}$ and the reduced dynamics of the system is obtained by tracing out the degrees of freedom of the environment, $\Dm^S(t)=\text{Tr}_E\left[\Dm(t)\right]=\text{Tr}_E\left[U(t)\Dm_0 U^\dagger(t)\right]$. This trace is conveniently evaluated in the orthonormal basis $\ens{\ket{\Enslabel}}$ of the environment,
 \begin{align*}
 		\Dm^S(t) &=  \int d\Enslabel p(\Enslabel) e^{-it(\Ham^S+ \Ham_\Enslabel)/\hbar} \Dm_0^S  e^{it(\Ham^S+ \Ham_\Enslabel)/\hbar} \\
 		&= \int d\Enslabel p(\Enslabel) U_\Enslabel(t) \Dm_0^S U_\Enslabel^\dagger(t) = \Dmap{\Dm_0^S}. \numberthis \label{eq:TrE}
 \end{align*}
Note that in practice, one first computes the $\textrm{Tr}_E$ using a sum $\sum_\Enslabel$ instead of the integral $\int d\Enslabel$, and subsequently takes the continuum limit of the sum to obtain \eqref{eq:TrE}. According to Eq.~\eqref{eq:TrE}, the reduced system dynamics forms a random unitary channel in the sense of Eq.~\eqref{eq:Rndunit Dynamical map}. 

Thus, we have shown how to obtain random unitary dynamics form the couplings of an open system to a static environment. We remark that the connection to disordered quantum systems is established by choosing the system Hamiltonian to be equal to the disorder average Hamiltonian $\Ham^S = \Avgham$, and the disorder realization Hamiltonian to be arising from the coupling to the environment, $\Ham_\Enslabel = \delta\Ham_\Enslabel$. Hence, the dynamics described by Eq.~\eqref{eq:Rndunit Dynamical map} can equivalently be derived from a microscopic disorder model or from a model of an open system coupled to an environment, or be postulated as a random unitary map (c.f. Fig.~\ref{fig:Equivalence illustration}).  

~\linebreak
~
Further on, we exploit the open systems perspective to derive a quantum master equation for the random unitary dynamics (\ref{eq:Rndunit Dynamical map}). 
Our derivation focuses on the weak-coupling limit and is analogous to the one presented, e.g., in \cite{Breuer2002}.

At the outset, we transform to the interaction picture, where operators evolve in time with $\Ham^S$, while states with $\HamI$. For clarity, we supply all interaction picture quantities with the subscript $I$. Next, it is useful to introduce the unitary time evolution operator, $\Uop_I = \Uop^\dagger_0(t)\Uop(t)$, where $\Uop_0(t) = e^{-it\Ham^S/\hbar} \otimes \Id$ and $\Uop(t)$ is defined above after Eq. (\ref{eq:H-Transfo}). Then, the state and the Hamiltonian in the interaction picture are given by $\Dm_I(t) = \Uop^\dagger_0(t) \Dm(t) \Uop_0(t)$ and $\Ham_I(t) = \Uop_0^\dagger(t) \HamI\Uop_0(t)$, respectively. Using Eq.  
\eqref{eq:Open syst Ham}, the latter Hamiltonian can be rewritten as,
\begin{align}
	\Ham_I(t) &=\int d\Enslabel e^{-it\Ham^S/\hbar}\Ham_\Enslabel e^{it\Ham^S/\hbar} \otimes \ketbra{\Enslabel}{\Enslabel}. \label{eq:BM interaction pic Ham}
\end{align}

In the interaction picture, the dynamics of a density operator obeys the von Neumann equation, 
\begin{align}\label{eq:BM interaction Von Neumann}
	\frac{d}{dt}\Dm_I(t) = -\frac{i}{\hbar}\commutator{\Ham_I(t)}{\Dm_I(t)}.
\end{align}
Equation \eqref{eq:BM interaction Von Neumann} is a typical starting point for the microscopic derivation of a master equation describing the dynamics of the reduced system, which here subsumes the random unitary channel. We proceed by integrating both sides of Eq.~\eqref{eq:BM interaction Von Neumann} over time, and subsequently inserting the result for $\Dm_I(t)$ back into Eq.~\eqref{eq:BM interaction Von Neumann}, to get
\begin{align}\label{eq:ST Dyson series}
	\dot{\Dm}_I(t) &= -\frac{i}{\hbar}\commutator{\Ham_I(t)}{\Dm(0)}\nonumber\\
	& -\frac{1}{\hbar^2}\int_0^t dt' \commutator{\Ham_I(t)}{\commutator{\Ham_I(t')}{\Dm_I(t')}}.
\end{align}
Note that no approximation has been made so far to derive the above integral-differential equation. It can be solved iteratively, yielding an infinite series in powers of $\Ham_I(t)$. Instead of doing so, let us take the trace over the environment degrees of freedom ($\textrm{Tr}_E$) in the exact Eq.~(\ref{eq:ST Dyson series}),
\begin{align}
	\dot{\Dm}_I^S(t) =& -\frac{i}{\hbar} \int d\Enslabel p(\Enslabel) \commutator{\Uop_0^{S \dagger}(t)\Ham_\Enslabel\Uop_0^S(t)}{\Dm^S_0} \label{eq:BM Dyson series}\\
	&-\frac{1}{\hbar^2} \textrm{Tr}_E\left\{\int_0^t dt' \commutator{\Ham_I(t)}{\commutator{\Ham_I(t')}{\Dm_I(t')}}\right\}, \nonumber
\end{align}
with $\Uop^S_0(t)=e^{-it\Ham^S/\hbar}$ the interaction picture unitary evolution operator of the system. The first term on the right-hand side of Eq.~\eqref{eq:BM Dyson series} vanishes due to the transformation \eqref{eq:H-Transfo}, which ensures that $\int d\Enslabel p(\Enslabel) \Ham_\Enslabel = 0$. As for the second term, we recall that in the Born approximation that we are using , $\Dm(t) = \Dm^S(t)\otimes \Dm^E$. This leads to the result,
\begin{align}
	\Dm_I(t')&=\Dm_I^S(t')\otimes\Dm^E  \nonumber \\&=\Uop_0^{S\dagger}(t')\Dm_0^S\Uop_0^S(t') \otimes \int d\Enslabel  p(\Enslabel) \ketbra{\Enslabel}{\Enslabel}. \label{eq:BM Born approx}
\end{align}
Inserting the latter result into Eq. (\ref{eq:BM Dyson series}) and performing the trace over the environment, we obtain
\begin{align*}
	\dot{\Dm}_I^S(t) =-\frac{1}{\hbar^2} \int_0^t dt' \int d\Enslabel p(\Enslabel) \commutator{{\Ham_\Enslabel}_I(t)}{\commutator{{\Ham_\Enslabel}_I(t')}{\Dm^S_I(t')}}, \numberthis \label{eq:BM interaction pic incoherent}
\end{align*}
where $\Ham_{\Enslabel I}(t) = U_0^\dagger(t) \Ham_\Enslabel U_0(t) $. 

We now make a further approximation, 
\begin{align} \label{eq:Markox approx}
	\Dm_I^S(t') \approx \Dm_I^S(t),
\end{align}
which renders Eq.~\eqref{eq:BM interaction pic incoherent} local in time. This approximation is justified in the regime of weak system-environment coupling and amounts to the truncation of the infinite series expansion by the terms that are second-order in $\Ham_I(t)$ \cite{Ishizaki2008}. In the context of disordered systems, this is equivalent to considering at most second-order correlations of the disorder potential.

Applying the approximation \eqref{eq:Markox approx} in Eq.~\eqref{eq:BM interaction pic incoherent}, we arrive at a Redfield master equation \cite{Breuer2002,Redfield1957} in the interaction picture,
\begin{align}
		\dot{\Dm}_I^S(t)&=-\frac{1}{\hbar^2} \int_0^t dt' \int d\Enslabel p(\Enslabel) \commutator{{\Ham_\Enslabel}_I(t)}{\commutator{{\Ham_\Enslabel}_I(t')}{\Dm^S_I(t)}}.
\end{align}
Dropping the (now, redundant) superscript $S$ of the density matrix and reverting back to the Schr\"odinger picture, we obtain the master equation,
\begin{align}\label{eq:BM QME}
	\dot{\Dm}(t) =-&\frac{i}{\hbar} \commutator{\Ham^S}{\Dm(t)} -\frac{1}{\hbar^2}\int d\Enslabel \,p_\Enslabel \commutator{\Ham_\Enslabel}{\commutator{\tilde{\Ham}_\Enslabel(t)}{\Dm(t)}},
\end{align}
where
\begin{align}\label{eq:BM interaction delta H}
	\tilde{\Ham}_\Enslabel(t) &\eqdef  \int_0^t dt' e^{-it'\Ham^S/\hbar}\Ham_\Enslabel e^{it'\Ham^S/\hbar}.
\end{align}
In the following, we explore the properties of Eq.~\eqref{eq:BM QME} for three different scenarios.

\textbf{a) Commuting system and interaction Hamiltonians:} 
As a first example, let us consider a system whose Hamiltonian commutes with the interaction Hamiltonian,
\begin{align}\label{eq:BM short-time approx condition}
\commutator{\Ham^S}{\Ham_\Enslabel}=0 \;,\; \forall \Enslabel.
\end{align}
In this case, Eq.~\eqref{eq:BM interaction delta H} is elementarily integrated to yield the relation
\begin{align}
	\tilde{\Ham}_\Enslabel(t) = t \Ham_\Enslabel.\label{eq:BM short time result}
\end{align}
Consequently, the master equation \eqref{eq:BM QME} becomes, 
\begin{align}\label{eq:BM Dephasing QME}
	\hspace{-0.34cm}\dot{\Dm}(t) =& -\frac{i}{\hbar}\commutator{\Ham^S}{\Dm(t)} -\frac{t}{\hbar^2}\int d\Enslabel p_\Enslabel \commutator{\Ham_\Enslabel}{\commutator{\Ham_\Enslabel}{\Dm(t)}}.
\end{align}
Equation (\ref{eq:BM Dephasing QME}) is easily solved in the eigenbasis $\ket{n}$ that is common to $\Ham^S$ and $\Ham_\Enslabel$, with the matrix elements of the density operator given by
\begin{align}\label{eq:matrix elements}
	\rho_{nm}(t)= \rho_{nm}(0) e^{-it(E_n^S-E_m^S)/\hbar }e^{- t^2C_2(n,m)/2\hbar^2},
\end{align}
where 
\begin{align}
	C_2(n,m) = \int d\Enslabel p_\Enslabel (E_n^{\Enslabel}-E_m^{\Enslabel})^2,
\end{align}
is the two-point correlation function which has the meaning of the average square of the energy gap between levels $n$ and $m$.
Equation (\ref{eq:matrix elements}) signifies pure dephasing dynamics: the diagonal elements ($n=m$) are time-independent, while the off-diagonal elements ($n\neq m$) evolve coherently with $\Ham^S$, and on top of that undergo a Gaussian decay with a linearly increasing in time rate $\gamma_{nm}(t) = t C_2(n,m)/\hbar^2$.  As it includes all contributions from the two-point correlations, the master equation \eqref{eq:BM Dephasing QME} is exact in the limit where the initial state distribution $p(\Enslabel)$ is a Gaussian. A rigorous proof thereof for finite-dimensional systems can be  easily derived by adapting corollary 3.1 of \cite{Wissmann2016}. 

Furthermore, taking the disorder instead of open system perspective, the condition \eqref{eq:BM short-time approx condition} is equivalent to the requirement that the disorder affects only the eigenvalues (i.e.,  spectral disorder) \cite{Kropf2016a}. On can then show that the exact dynamics also results in pure dephasing \cite{Kropf2016a}, and, hence, is not a consequence of the weak-coupling approximation given by Eq.~\eqref{eq:Markox approx}, but is a direct consequence of the commutation relation \eqref{eq:BM short-time approx condition}.

\textbf{b) Short times:} Let us now study the behaviour of $\Dm(t)$ at short times. For that purpose, we expand Eq.~\eqref{eq:BM interaction delta H} to first order in time, which yields the same result as for condition \eqref{eq:BM short-time approx condition}, namely $\tilde{\Ham}_\Enslabel(t) = t  \Ham_\Enslabel$. Hence, again, the corresponding master equation is Eq.~\eqref{eq:BM Dephasing QME}. Note that, since the Hamiltonians $\Ham_\Enslabel$ for different $\Enslabel$ may not commute with each other, in this case the dynamics does not necessarily result in pure dephasing. Comparing the first- and second-order terms in the series expansion of the exponential in Eq.~\eqref{eq:matrix elements}, the validity time $\tau$ can be estimated to be proportional to the mean level spacing of the system Hamiltonian,
\begin{align}\label{eq:ST general validity}
	\tau  \approx \frac{\hbar}{\left<E_m^S-E_n^S\right>}.
\end{align}
This time scale is given by the energy-time uncertainty relation \cite{LeBellac2006}, and is sometimes termed Heisenberg time (see, e.g., \cite{Wimberger2014}). 

\textbf{c) Markov approximation:} A common approximation in the study of open quantum systems is the semi-group Markov approximation which leads to the Gorini-Kossakowski-Sudarshan-Lindblad (GKSL) master equation \cite{Lindblad1976,Gorini1976}. To derive the latter, we begin by expanding Eq.~\eqref{eq:BM interaction delta H} in the eigenbasis $\ket{n}$ of the system Hamiltonian
\begin{align}
	\hspace{-0.2cm}\tilde{\Ham}_\Enslabel(t) =& \sum_{m,n=1}^\Dim \ketbra{m}{m}\Ham_\Enslabel\ketbra{n}{n} \int_0^t dt' e^{-\frac{i}{\hbar}(t-t')(E^S_m-E^S_n)},  \label{eq:BM delta H expansion}
\end{align}
with $E^S_n$ the $n^{th}$ eigenvalue of $\Ham^S$ and $d={\rm dim}( \Ham^S)$. 

The semigroup Markov approximation is obtained by setting the upper integration limit of the time integral in Eq.~\eqref{eq:BM delta H expansion} to infinity:
\begin{align} \label{eq:BM Markov limit}
\limit{t}{\infty}\int_0^t dt' e^{-\frac{i}{\hbar}(t-t')(E^S_m-E^S_n)} = \limit{\epsilon}{0}\frac{i\hbar}{E^S_n-E^S_m+i\epsilon}=: R_{mn}.
\end{align} 
We then obtain
\begin{align} \label{eq:Ham Markov}
	\tilde{\Ham}_\Enslabel = \sum_{m,n=1}^\Dim R_{mn} \matelem{m}{\Ham_\Enslabel}{n}\ketbra{m}{n}.
\end{align}
The above expression \eqref{eq:BM Markov limit} is the matrix element $R_{mn}=\matelem{m}{\op{G}_S(E_n)}{m}$ of the free retarded resolvent operator $\op{G}_S(E)=(E-\Ham^S+i\epsilon)^{-1}$. In other words, $R_{mn}$ is the transition amplitude from state $\ket{m}$ to $\ket{n}$ of a particle evolving only with the system Hamiltonian $\Ham^S$. 

Inserting Eq.~\eqref{eq:Ham Markov} into Eq.~\eqref{eq:BM QME} yields a GKSL master equation, which is by definition time-independent. Hence, it is clear that the Markov approximation cannot be applied to the previous two examples, which are characterized by the time-dependent master equation \eqref{eq:BM Dephasing QME}.
 Finally, we note that when the GKLS master equation arises as a consequence of averaging over random impurities in a disordered sample, one obtains pure momentum-dephasing random unitary dynamics \cite{Muller2009}.

~\linebreak

In this article we studied a particular class of random unitary channels and showed how to embed the latter in the framework of the theory of open quantum systems. Making use of the fact that our random unitary channels imply the Born approximation, we derived the corresponding Redfield master equation. We then studied in more detail three specific cases: commuting system and interaction Hamiltonians, the short-time approximation and the semi-group Markov approximation. We showed that in the first case, one obtains a pure dephasing master equation with rates that are linearly increasing with time. In the second case, we obtained a master equation with the same rates, but that does not invariably lead to dephasing dynamics. In the semi-group Markov approximation, the resulting GKLS master equation is characterized by the transition amplitudes generated by the system Hamiltonian. 

In conclusion, the dynamics of quantum systems subject to classical uncertainty can be studied either starting from the random unitary Kraus map Eq.\eqref{eq:Rnd unit map general} \cite{Bengtsson2006}, from an ensemble of disordered Hamiltonians \cite{Kropf2016a}, or, as discussed here, from a system coupled to a static environment. Each approach allows for specific computational methods and provides particular grounds for experimental implementations. We believe that these different points of view are fruitful for fostering synergies between the fields of open quantum systems, disorder physics and information theory \cite{Kropf2017}.\hfill \\

The authors wish to express their gratitude for discussions with Clemens Gneiting. C. K. acknowledges funding by the German National Academic Foundation.


\appendix


\bibliographystyle{apsrev4-1}
\bibliography{bibliography_MicroRndUnit}

\end{document}